\documentclass[preprint,epsfig,aps,prb]{revtex4}
\usepackage{graphicx}
\begin{document}

\title{Polarization control of optical transmission of a periodic array of elliptical holes in a
metal film}

\author{Jill Elliott,$^1$ Igor I. Smolyaninov,$^2$ Nikolay I. Zheludev,$^3$ Anatoly V. Zayats$^1$\footnote{Author to whom correspondence should be addressed:
electronic mail: a.zayats@qub.ac.uk.}}
\address{$^1$School of Mathematics and Physics, The Queen's
University of Belfast, Belfast BT7 1NN, UK}

\address{$^2$Electrical and
Computer Engineering Department, University of Maryland, College
Park, MD 20742, USA}

\address{$^3$School of Physics and Astronomy, University of Southampton, Southampton SO17 1BJ,
UK}

\date{\today}
%\maketitle
\begin{abstract}
Spectral dependencies of polarized optical transmission of a metal
film with a periodic array of elliptical nanoholes have been
studied. Such nanostructured metal films exhibit the enhanced
broadband optical transmission which can be controlled by
selecting polarization of incident and/or transmitted light.
\end{abstract}
\maketitle

Optical components for applications in extreme ambient conditions
and high level of integration require novel design approaches.
Recent progress in understanding of the optical properties of
nanostructured metal films has allowed the design of spectral
filters and devices which exploit surface plasmon polariton
excitations (SPPs) in metal nanostructures.\cite{ebbessen-98,jopa}
Such nanostructures could act as band-pass spectral
filters\cite{ebbessen-98,krishnan-01} or, if combined with
nonlinear materials, could perform all-optical stitching and
limiting functions at low control light
intensities.\cite{prl-02,apl-02,prb-02,dykhne-03} Polarisation
properties of light can also be controlled with metallic
nanostructures.\cite{zheludev-03,sanders}

Functionality of metallic nanostructures are determined by the
interaction of photons with collective excitations of conduction
electrons at a metal surface.\cite{agran} The coupling and mutual
transformations between photons and surface plasmons depend upon
the frequencies and wave vectors of the excitations, as well as on
the polarization state of incident electromagnetic field. The
polarization state determines the excitation of SPPs in one or
another generally non-equivalent directions of the Brillouin zone
related to the periodic structure. By choosing the light
polarization one could not only provide the optimal intensity
transmittance through the structure but could also introduce
various polarization-sensitive effects related to the chirality of
the structure and non-locality of its electromagnetic response.
\cite{zheludev-03,schwanecke-03}

In this letter we report the studies of polarization properties of
the enhanced light transmission through a metal film with a
periodic arrangement of oriented elliptical nanoholes. Such
metallic structure exhibits a broadband transmission stemming from
the multiple SPP resonances permitted by the low-symmetry
arrangement of elliptical holes in a square array. The
transmission spectrum strongly depends on the polarization state
of the incident light even at normal incidence, the property of
striking contrast with a square array of circular holes where
transmission spectrum has well-defined resonances independent on
the polarization of the incident light.\cite{prl-01} In the case
of elliptical holes, the incident linearly polarized light is
converted into elliptically polarized in transmission. Using this
feature we can continuously tune the transmission spectrum of the
structure from blue to red wavelengths by changing the
polarization state of the incident light and selecting appropriate
polarization component of the transmitted light.

\begin{figure}
\begin{center}
\includegraphics[width=3.5in]{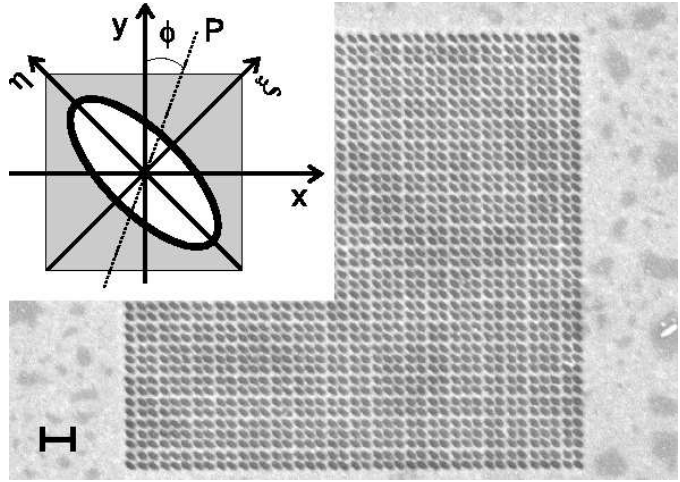}
\end{center}
\caption{The image of the periodic array of elliptical holes in a
gold film. Scale bar is 2 $\mu$m. Schematic of the primitive cell
of the lattice is shown in the inset.}\label{structure}
\end{figure}

The metallic nanostructures of overall size of $20\times 20$
$\mu^2$ studied here were fabricated in a 40 nm gold film on a
glass substrate using ion beam milling. The structure consists of
elliptical holes with the main axes of the ellipse of 250 and 500
nm (Fig. \ref{structure}). The holes are arranged in a square
array with period $D=$500 nm. The transmission spectra of the
nanostructures were recorded using a spectrograph equipped with a
CCD-camera coupled to a long-working-distance optical microscope.
Collimated white light from a stabilized tungsten-halogen source
passed through a polarizer and illuminated the sample at normal
incidence. The light transmitted through the nanostructure was
collected by an objective lens. The beam position at the
nanostructure was monitored by an imaging CCD camera. The light
exiting the structure passed through the analyzer and was launched
into a fiber bundle connected to the spectrograph. Polarization
mode scrambling in the fiber bundle ensured that
polarization-related effects do not influence spectral
measurements.

\begin{figure}
\begin{center}
\includegraphics{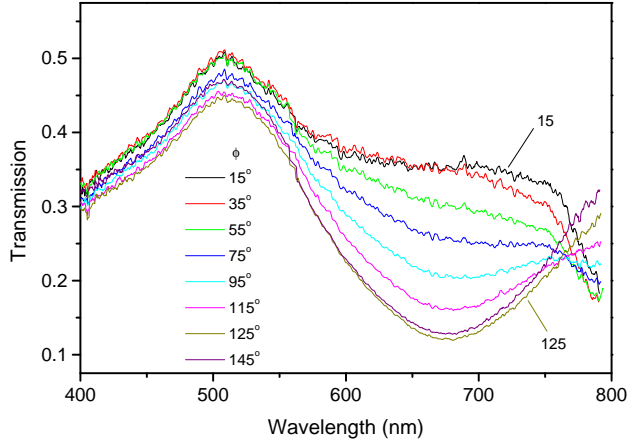}
\end{center}
\caption{Normal incidence transmission spectra of the array of
elliptical holes for different polarizations of the incident light
without polarization analysis of the transmitted light.
Polarization angle $\phi$ is measured with respect to $y$-axis
(Fig. \ref{structure}).}\label{unpol}
\end{figure}

The transmission spectra of the structure obtained at normal
incidence without polarization selectivity of the transmitted
light are shown in Fig. \ref{unpol} for different polarization
states of the incident light with respect to the array
orientation. For the incident light polarization parallel to the
principal lattice axes ($x$ or $y$) the transmission spectrum is
observed with broad features in the 450--550 nm and 750--800 nm
spectral ranges. With the increase of the polarization azimuth
$\phi$ up to about 20--25$^o$ the transmission increases in the
550--750 nm wavelength range, where the spectrum becomes rather
flat. With further increase of $\phi$ when the light is polarized
along the short principal axis of the elliptical holes, the
transmission in the long-wavelength spectral range (750--800 nm)
is significantly suppressed. For the polarisation azimuth along
the long principal axis of ellipse, the transmission at the far
red part of the spectrum is enhanced, but becomes smaller at
around 600--700 nm. Thus, with rotation of the polarization of the
incident light, the intensity of the spectral components of the
transmitted light exhibit complex, oscillatory behavior different
in different spectral ranges. The relatively well-defined
long-wavelength band at around 780--800 nm has transmission
minimum for the polarization of light parallel to the short
principal axis of an ellipse and maximum for the polarization
along the long principal axis. For all wavelengths, the
dependencies show a 2-fold symmetry corresponding to the symmetry
of the array: a square lattice with basis elements of a 2-fold
rotational symmetry.
%(elliptical holes).
%For all wavelengths the dependencies
%show a 2-fold symmetry corresponding to the symmetry of the array:
%a square lattice with an elliptical basis.

To understand polarization sensitivity of transmission, the
spectrum of the SPP Bloch waves on a periodically structured film
should be considered. The strict consideration of the symmetry
properties of the eigenstates of a periodic structure with an
asymmetrical lattice basis would require the approach analogous to
the linear combination of atomic orbitals (LCAO) method or the
Wannier fanctions conventionally used in solid state
physics.\cite{kittel} Nevertheless, it will be instructive to
consider a simplified model using the SPP Bloch modes.

The field of SPP modes on a periodically structured surface can be
described by
\begin{equation}\label{bloch}
E(x,y) = U_{{\mathbf k}_{SP}} (\xi,\eta) exp[i (k^{(x)}_{SP}x +
k^{(y)}_{SP}y)] ,
\end{equation}
where $U_{\mathbf k_{SP}} (\xi,\eta)$ possesses the periodicity of
the array and is the SPP Bloch function and ${\mathbf k}_{SP}$ is
the wave vector of the Bloch wave. Here the coordinate systems
($\xi$,$\eta$) of the basis and ($x$,$y$) of the lattice are not
generally independent and introduced to simplify symmetry
considerations. At normal incidence of light, the allowed SPP
wavevectors on the periodically structured surface are determined
by
\begin{equation}
{\mathbf k}_{SP}= \pm p \frac{2\pi}{D}{\mathbf u}_x \pm q
\frac{2\pi}{D} {\mathbf u}_y , \label{k-states}
\end{equation}
where ${\mathbf u}_x$ and ${\mathbf u}_y$ are the unit reciprocal
lattice vectors of the periodic structure, $D$ is its periodicity
(assumed to be the same in both $x$- and $y$-directions), and $p$
and $q$ are integer numbers corresponding to the different
directions of the SPP Brillouin zone. At normal incidence, SPP can
be excited if the electric field of the incident light has a
component in the direction of SPP propagation: $({\mathbf E}\cdot
{\mathbf k}_{SP})\neq 0$. Thus, the polarization dependence of the
coupling efficiency in different directions of the Brillouin zone
is proportional to $|pcos\phi+qsin\phi|$, where $\phi$ is the
polarization azimuth angle with respect to the $x$ principal axis
of the lattice. It is clear from this analysis that for a square
lattice with basis elements of circular symmetry no polarization
dependencies of the SPP excitation and, thus, of the enhanced
transmission can be expected at normal incidence.

However, in the case of an elliptical basis of the lattice,
polarization dependencies are significant. Taking into account
different scattering properties of the ellipse in different
directions, one can understand this by considering the mixing of
the SPP Bloch states of the square lattice due to the lower
symmetry of the basis elements. Elliptical holes with different
size in the different directions modify the Brillouin zone
structure by introducing size-dependent anizotropic scattering.
This leads to the reduction of the Brillouin zone symmetry showing
a 2-fold rotation axis instead of 4-fold, appropriate for a square
lattice, effectively creating a medium with pronounced linear
birefringence in dichroism. Linearly polarized waves with
polarization azimuth along the main axes of the elliptical basis
will be the polarization eigenstates of the structure. As a result
of the anizotropic SPP Brillouin zone, the transmission spectrum
of the structure is dependent on the polarization of the incident
light.

\begin{figure}
\begin{center}
\includegraphics{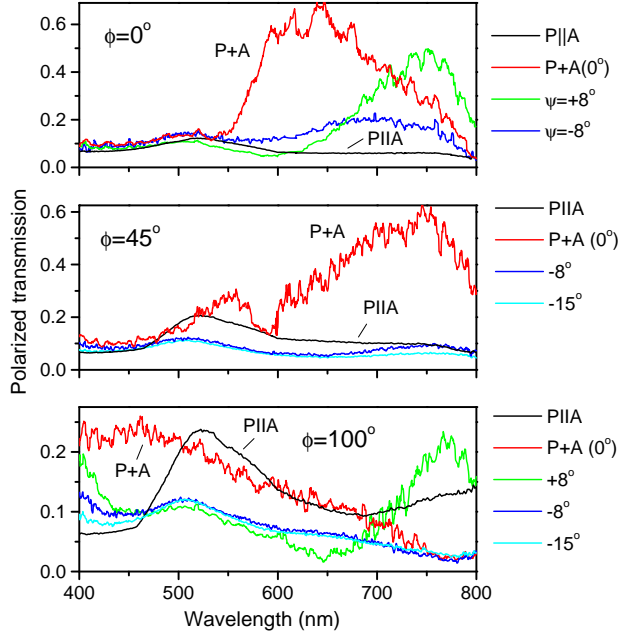}
\end{center}
\caption{Spectra of polarized transmission of the array measured
for the incident light polarization (a) $\phi=0^o$ and (b) 45$^o$,
(c) 100$^o$. The spectra of the light transmitted without
polarization state change ($P\parallel A$), with orthogonal
polarization ($P\perp A$, $\psi =0^o$), and close to orthogonal
polarization $\psi = \pm$8$^o$ and 15$^o$ are shown.}\label{depol}
\end{figure}

\begin{figure}
\begin{center}
\includegraphics{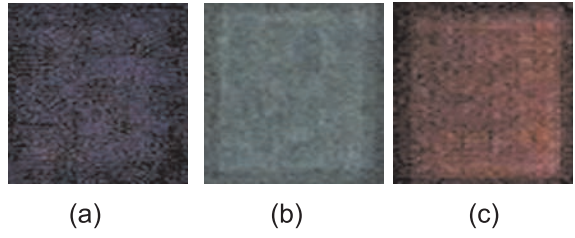}
\end{center}
\caption{True colour images of the nanostructure taken in
transmission for different polarizations of the incident and
transmitted light showing tuning of the dominating transmission
wavelength.}\label{images}
\end{figure}

Mixing and re-excitation of surface plasmon resonances in
different directions of the Brillouin zone are responsible for
anizotropic retardation and eventually for the elliptization of
the polarization state of the transmitted light. Polarized
transmission spectra show that the changes of the polarization
state of the transmission light are strongly wavelength dependent
(Fig. \ref{depol}). This allows identification of transmission
resonances contributing to the broad transmission spectrum.
Several resonances can be identified by the different degree of
elliptization at around 750, 650, and 550 nm. Thus, by controlling
the polarization state of the incident and exercising polarization
selection of the transmitted light, transmission spectrum of the
elliptical hole array can be tuned at specific resonant band
corresponding to one or another SPP mode of the system,
quasi-continuously from blue to red wavelengths of the visible
spectrum (Fig. \ref{images}).

In conclusion, we have studied polarization properties of the
transmission spectra of a periodic array of elliptical holes in a
metal film. Due to the reduced symmetry of the lattice, elliptical
holes allow to achieve the enhanced broadband transmission which
can be controlled by choosing the polarization of the incident
and/or transmitted light, offering numerous applications in
passive and active optical components. Combined with non-linear
effects which are significantly enhanced in metallic structures
due to surface plasmons, the observed polarization properties will
allow further applications in integrated photonic circuits.

\acknowledgements{We are indebt to C. Cassidi for sputtering thin
gold films used in these experiments. This work was supported in
part by EPSRC.}


\newpage



\begin{references}

\bibitem{ebbessen-98}W.L. Barnes, A. Dereux, T.W. Ebbesen, Nature,
\textbf{424}, 824 (2003).

%T. W. Ebbesesn, J. Lezec, H. F. Ghaemi, T. Thio, and P. A. Wolff, Nature (London) {\bf 391}, 667 (1998).

\bibitem{jopa}A. V. Zayats and I. I. Smolyaninov, J. Opt. A
\textbf{5}, S16 (2003).

\bibitem{krishnan-01}A. Krishnan, T. Thio, T. J. Kim, H. J. Lezec, T. W. Ebbesen, P.A.
Wolf, J. Pendry, L. Martin-Mareno, and F. J. Garsia-Vidal, Optics
Comm. \textbf{200}, 1 (2001).

\bibitem{prl-02}I. I. Smolyaninov, A. V. Zayats, A. Gungor, and C. C. Davis, Phys. Rev.
Lett. \textbf{88}, 187402 (2002).

\bibitem{apl-02}I. I. Smolyaninov, A. V. Zayats,
and C. C. Davis, Appl. Phys. Lett. {\bf 81}, 3314 (2002).

\bibitem{prb-02} I. I. Smolyaninov, A. V. Zayats, A. Stanishevsky, and C. C. Davis, Phys. Rev.
{\bf B66}, 205414 (2002).

\bibitem{dykhne-03}A.M. Dykhne, A.K.Sarychev, and V.M.Shalaev, Phys.Rev. B
\textbf{67} (2003) 195402.

\bibitem{zheludev-03}A. Papakostas, A. Potts, D. M. Bagnall, S. L. Prosvirnin,
H. J. Coles, N. I. Zheludev, Phys. Rev. Lett. {\bf 90}, 107404
(2003).

\bibitem{sanders}I. R. Hooper and J. R. Sambles,
Opt. Lett. {\bf 27} 2152 (2002).


\bibitem{agran}{\it Surface Polaritons}, V.
M. Agranovich and D. L. Mills, Eds., North-Holland, Amsterdam,
1982.

\bibitem{schwanecke-03}A. S. Schwanecke, A. Krasavin, D. M. Bagnall, A. Potts, A. V.
Zayats, N. I. Zheludev, Phys. Rev. Lett., {\bf 91}, NNNNN (2003).

\bibitem{darm-prb-03} S. A Darmanyan and A. V. Zayats, Phys. Rev. B.
67, 035424 (2003).

\bibitem{prl-01}L. Salomon, F. Grillot, F. de Fornel, and A. V. Zayats, Phys. Rev.
Lett. \textbf{86}, 1110 (2001).

\bibitem{kittel} C. Kittel, {\it Introduction to Solid State Physics}, Wiley,
Chichester, 1996.

\end{references}
\end{document}